\documentclass[preprint,amsmath,amssymb,aps,prc,floatfix,nofootinbib]{revtex4-2}

\usepackage[utf8]{inputenc}
\usepackage[T1]{fontenc}
\usepackage{graphicx}
\usepackage{amsmath}
\usepackage{mathtools}
\usepackage{hyperref}
\usepackage[inkscapelatex=false]{svg}
\usepackage{simplewick}
\usepackage{braket}
\usepackage{xcolor}

\usepackage{etoolbox}

\makeatletter
\patchcmd{\frontmatter@abstract@produce}
  {\vskip200\p@\@plus1fil
   \penalty-200\relax
   \vskip-200\p@\@plus-1fil}
  {}
  {}
  {}
\makeatother

\newcommand{\rr}{\boldsymbol{r}}
\newcommand{\q}{\boldsymbol{q}}
\newcommand{\kk}{\boldsymbol{k}}
\newcommand{\p}{\boldsymbol{p}}
\newcommand{\PP}{\boldsymbol{P}}
\newcommand{\oq}{\Omega_{\q}}
\newcommand{\idoq}{\int d\Omega_{\q}}
\newcommand{\idoqa}{\frac{1}{4\pi} \idoq}
\newcommand{\idor}{\int d\Omega_{\rr}}
\newcommand{\ei}{\varepsilon^i}
\newcommand{\ej}{\varepsilon^j}
\newcommand{\eij}{\varepsilon^{ij}}
\newcommand{\mage}{|\boldsymbol{\varepsilon}|^2}
\newcommand{\zzeta}{\boldsymbol{\zeta}}
\newcommand{\kbar}[1]{\bar{K}_{#1}}

\newcommand{\sbar}[1]{\bar{S}_{#1}}
\newcommand{\dbar}{\bar{D}}
\newcommand{\ksbar}[1]{\kbar{#1}\sbar{#1}}

\newcommand{\rhat}{\hat{\rr}}
\newcommand{\qhat}{\hat{\q}}
\newcommand{\xhat}{\hat{\boldsymbol{x}}}

\newcommand{\zhat}{\hat{\boldsymbol{z}}}

\newcommand{\tre}{\mathrm{tr}(\varepsilon)}

\newcommand{\oo}[1]{\mathcal{O}(#1)}
\newcommand{\Cs}{C_\mathrm{smooth}}

\newcommand{\Ca}{C_\mathrm{aniso}}
\newcommand{\B}{\mathcal{B}}

\newcommand{\e}{\varepsilon}

\begin{document}

\title{Corrections to the Smoothness and On-Shell Approximations in Femtoscopy and Coalescence}
\author{Isaac G. Smith}
\email[email: ]{isaacsmi@weizmann.ac.il}
\affiliation{Weizmann Institute of Science, Rehovot 7610001, Israel}

\author{Kfir Blum}
\email[email: ]{kfir.blum@weizmann.ac.il}
\affiliation{Weizmann Institute of Science, Rehovot 7610001, Israel}

\begin{abstract}
Relativistic heavy-ion collisions produce femtometer-scale sources whose space-time structure can be constrained using two-particle femtoscopic correlations. Standard implementations rely on the smoothness and on-shell approximations, which effectively remove the relative momentum dependence of the particle emission function. We explore the validity of these approximations by deriving model-independent expansions that quantify the leading corrections for femtoscopy and coalescence with arbitrary sources and final-state interactions. The resulting first- and second-order correction terms can be evaluated with essentially the same numerical complexity as the usual Koonin-Pratt expressions; for angle-averaged correlations the first-order contributions vanish by symmetry.
We illustrate the framework with explicit calculations in a blast-wave source model; for blast-wave parameter sets representative of pp and PbPb fits at LHC energies, the corrections are at the few percent level for pp correlations and deuteron coalescence. These corrections are potentially subdominant compared to other effects, for example corrections to the equal time approximation.

This version of our paper corrects a numerical error in a previous version. We find that the smoothness corrections are numerically larger than previously reported.
\end{abstract}
\maketitle

\section{Introduction} \label{sec1}
Relativistic heavy-ion collisions at RHIC and LHC energies produce a high-excitation state (HXS), the properties of which are important for high-energy physics\cite{Shuryak:2004cy,Citron:2018lsq,Busza:2018rrf} and astrophysics\cite{Blum:2017qnn}. 
The asymptotic momenta of particles produced in the collision are correlated by the dynamics of the HXS and by final-state interactions (FSI). 
Measuring these correlations allows for the determination of spatial and temporal properties of the HXS, via a method known as femtoscopy\cite{lisa-review,heinz-review,weidemann-heinz,bellini-blum}. Additionally, 
there have been attempts to determine the parameters of FSI potentials by 
comparing the correlations of different species under the assumption of a universal underlying particle source model, with an application for hyperon interactions\cite{alice-strange1,alice-strange2,alice-strange3,alice-strange4,nstar}\footnote{See \cite{epelbaum} for a critical assessment of the definition of a universal particle source.}. The particle source inferred from femtoscopy can also be used to predict  bound-state coalescence (see, e.g., \cite{Mrowczynski:1987oid,bwm,finite-size,Mrowczynski:2016xqm}), a result known as the coalescence-correlation relation\cite{blum-takimoto,bellini-blum}. 

Several approximations are often employed in femtoscopy and coalescence analyses. We study two of these approximations which are closely related, termed the smoothness and on-shell approximations, aiming to quantify the error induced by adopting these approximations. We derive model-independent first- and second-order corrections, that can be evaluated with the same level of complexity as the leading-order expressions (though we show that in the case of angle-averaged correlations the first-order corrections vanish due to symmetry considerations). 
This provides a practical tool for assessing the validity of the approximations for any given source model. 
Others have assessed these approximations' validity\cite{pratt-smoothness} and derived corrections to the approximations\cite{alt-smoothness1} both in the specific case of free identical particle correlations, and a less-strict smoothness approximation has also been proposed\cite{alt-smoothness2}. The main advantage of our method, however, is the ability to compute corrections to the smoothness and on-shell approximations for general sources in the presence of nontrivial FSI; we have not found such a tool in existing references.

This work is organized as follows. In Sec. \ref{sec2} we provide a brief overview of the calculation of correlations and coalescence within the smoothness approximation, including a discussion of other approximations which we do not investigate. In Sec. \ref{sec3} we present the corrections to the smoothness approximation, which we call the smoothness expansion. We begin with the expansion for angle-dependent correlations (Sec. \ref{sec3a}) and then move on to angle-averaged correlations (Sec. \ref{sec3b}, with additional details in App. \ref{appA}). We then work out the explicit corrections to the correlations of free particles (Sec. \ref{sec3c}) before applying the smoothness expansion to coalescence (Sec. \ref{sec3d}). Next, we derive the analogous corrections for the on-shell approximation in Sec. \ref{sec4}. In Sec. \ref{sec5} we perform explicit numerical calculations within the phenomenological blast wave model (explained in Sec. \ref{sec5a} with additional details in App. \ref{appB}) for proton-proton correlations (Sec. \ref{sec5b}) and deuteron coalescence (Sec. \ref{sec5c}). We show that for this particular model, the corrections are comparable to state-of-the-art experimental uncertainties for parameters representative of pp collisions. We conclude in Sec. \ref{sec6}.

\section{Theoretical Background} \label{sec2}
Consider the number $N_{2,s}$ of pairs produced with constituent four-momenta $p_1^\mu$ and $p_2^\mu$ in a given spin channel $s$. By mixing pairs from different events, one can also construct the number $N_2^0$ of these pairs which would be produced if FSI and particle statistics were turned off (up to effects from non-femtoscopic correlations)\cite{lisa-review}. The femtoscopic correlation function is then defined as (see, e.g. \cite{bellini-blum})
\begin{equation}
    C(p_1^\mu,p_2^\mu) = \frac{\sum_s p^0_1 p^0_2 \frac{dN_{2,s}}{d^3\p_1d^3\p_2}}{p^0_1 p^0_2 \frac{N_2^0}{d^3\p_1d^3\p_2}}.
\end{equation}
We make the approximation that after some freeze-out time the particles' dynamics are dominated by their interaction with each other (ignoring all other particles). This is called the sudden approximation. The pair can then be modeled as being produced at freeze-out via a two-particle source function $S(r^\mu,k^\mu;p^\mu)$ (the arguments of which will be explained below) that encompasses all interactions prior\cite{bellini-blum}.

Now, for particles with respective masses $m_1$ and $m_2$ we write
\begin{equation}\label{mom-decomp}
    p_1^\mu = (1+\alpha)P^\mu + q^\mu, \qquad p_2^\mu = (1-\alpha)P^\mu - q^\mu
\end{equation}
where
\begin{equation}\label{alpha}
    \alpha = \frac{m_1^2 - m_2^2}{4P^2}.
\end{equation}
For equal masses $\alpha=0$ so we get $q^\mu=\frac{p_1^\mu-p_2^\mu}{2}$. Note that there exist other conventions (for example in \cite{alt-smoothness1}) where $q^\mu$ is defined with an extra factor of 2, i.e. $q^\mu=p_1^\mu-p_2^\mu$ in the equal mass case.

For most of our analysis it will be convenient (and conventional) to express the kinematical variable $q^\mu$ in the pair rest frame (PRF), where it is purely spatial: $q^\mu=(0,{\q})$. In contrast, the average pair momentum $P^\mu$ is conveniently expressed in the lab frame (note that in the PRF, $P^\mu$ is purely temporal). 
%
The correlation function becomes\cite{bellini-blum}
\begin{equation}\label{cqp_rel}
    C(\q;P^\mu) = \sum_s w_s \frac{\int d^4 r \int \frac{d^4k}{(2\pi)^4} \,D_{\q,s}\left(r^\mu,k^\mu\right) S(r^\mu,k^\mu;P^\mu)}{\int d^4 r \,S(r^\mu,q^\mu;P^\mu)\mid_{q^0=0}},
\end{equation}
where
\begin{equation}\label{dq}
    D_{\q,s}\left(r^\mu,k^\mu\right) = \int d^4\zeta e^{ik_\mu \zeta^\mu} \phi_{\q,s}\left(r^\mu+\frac{\zeta^\mu}{2}\right) \phi_{\q,s}^* \left(r^\mu-\frac{\zeta^\mu}{2}\right)
\end{equation}
is the relativistic Wigner density for the scattering Bethe-Salpeter amplitude $\phi_{\q,s}(r^\mu)$ with asymptotic outgoing momentum $\q$; and $w_s$ is the weight of the spin channel. In Eq.~(\ref{cqp_rel}), $r^\mu$ and $k^\mu$ are the relative phase space coordinates for the pair, which we also define in the PRF. Here and in what follows, when PRF quantities (like $\q$) appear alongside lab-frame quantities (like $P^\mu$) as the arguments of a function (like $C(\q;P^\mu)$), we keep the lab-frame argument behind semicolon.


We now invoke the equal-time approximation (ETA)\cite{finite-size}, which states that the Bethe-Salpeter amplitude is approximately independent of time, and reduces to the Schrodinger wavefunction $\phi_{\q,s}(\rr)$. Then the relativistic Wigner density can be related to the nonrelativistic Wigner density $D_{\q,s}(\rr,\kk)$ as
\begin{equation}
    D_{\q,s}\left(r^\mu,k^\mu\right) \approx 2\pi \delta(k^0) D_{\q,s}(\rr,\kk) = 2\pi \delta(k^0) \int d^3\zzeta e^{-i\kk\cdot \zzeta} \phi_{\q,s}\left(\rr+\frac{\zzeta}{2}\right) \phi_{\q,s}^* \left(\rr-\frac{\zzeta}{2}\right),
\end{equation}
which lets us write the correlation function within the ETA as
\begin{equation}\label{cqp}
    C(\q;P^\mu) = \sum_s w_s \frac{\int d^3 \rr \int \frac{d^3\kk}{(2\pi)^3} \,D_{\q,s}\left(\rr,\kk\right) S(\rr,\kk;P^\mu)}{\int d^3 \rr \,S(\rr,\q;P^\mu)}
\end{equation}
using the equal-time source function
\begin{equation}
    S(\rr,\kk;P^\mu) = \int dr^0 S(r^\mu,k^\mu;P^\mu)\mid_{k^0 = 0}.
\end{equation}

The traditional next step is to apply two closely-related approximations\cite{weidemann-heinz}. Together, they amount to the source function being independent of the relative momentum. First is the on-shell approximation, which arises from a subtlety involved with the average CM momentum $P^\mu$, as it is in general $\q$-dependent. This also means that the definition of the PRF is dependent on $\q$. We define a momentum $p^\mu$ which is the average CM momentum when $\q=0$, i.e.
\begin{equation} \label{p-vs-P}
    \p\equiv\PP,\qquad p^2 = \left(\frac{m_1+m_2}{2}\right)^2\qquad \Rightarrow \qquad p^\mu = \left(\sqrt{\left(\frac{m_1+m_2}{2}\right)^2 + \PP^2}, \PP\right).
\end{equation}
We approximate the PRF as the rest frame of $p^\mu$ for all $\q$ (equivalently, we approximate $P^\mu\approx p^\mu$). Notably, $(1\pm \alpha_0)p^\mu$ is on-shell with mass $m_1$ ($+$) or $m_2$ ($-$) where
\begin{equation}\label{alpha0}
    \alpha_0 = \frac{m_1^2-m_2^2}{(m_1+m_2)^2} = \frac{m_1-m_2}{m_1+m_2}
\end{equation}
is $\alpha$ at $\q=0$. This means that the rest frame of $p^\mu$ is the rest frame of physical particles with momenta $(1\pm \alpha_0)p^\mu$, while the same does not hold for $(1\pm \alpha)P^\mu$ unless $\q=0$ (in which case $P^\mu =p^\mu$ and $\alpha=\alpha_0$). Thus, our approximation of $P^\mu\approx p^\mu$ is termed the on-shell approximation. For particles with different masses this approximation also involves $\alpha\approx \alpha_0$, while for equal masses $\alpha=\alpha_0=0$ exactly for all $\q$. The benefit of this approximation becomes more clear when in conjunction with the next approximation, the smoothness approximation.

The smoothness approximation\cite{original-smoothness} states
\begin{equation}\label{smoothness}
    S(\rr,\q;p^\mu) \approx S(\rr,0;p^\mu)\equiv S(\rr;p^\mu).
\end{equation}
If we normalize $S(\rr;p^\mu)$ such that it integrates to 1 over all space, noting the identity $\int \frac{d^3\kk}{(2\pi)^3} \,D_{\q,s}\left(\rr,\kk\right) = \left| \phi_{\q,s}(\rr) \right|^2$ we get the Koonin-Pratt formula\cite{koonin-kp,pratt-kp}
\begin{equation}\label{kp}
    C(\q;p^\mu) \approx \sum_s w_s \int d^3 \rr \left| \phi_{\q,s}(\rr) \right|^2 S(\rr;p^\mu).
\end{equation}
Applying the smoothness and on-shell approximations together ensures that we do not evaluate the source function at off-shell momenta, which allows for the use of source models which are only defined on-shell (e.g. from classical hydrodynamic or kinetic theory simulations). However, in certain cases the approximations create pathological behavior, such as free bosonic correlation functions dropping below unity\cite{inconsistency,pratt-smoothness}. In this work, we will investigate corrections to both of these approximations.

Before turning to our corrections, we also note the analogous formalism for coalescence. The coalescence factor for particles with four-momenta $p_1^\mu$ and $p_2^\mu$ (masses $m_1$ and $m_2$, spins $s_1$ and $s_2$) coalescing into a particle with four-momentum $p_c^\mu$ (mass $m_c$, spin $s_c$, with $N_c$ particles produced) is defined as
\begin{equation}
    \B = \frac{p^0_c \frac{dN_c}{d^3\p_c}}{p^0_1 p^0_2 \frac{N_2^0}{d^3\p_1d^3\p_2}},
\end{equation}
where this is evaluated when $p_c^\mu$ is at rest in the PRF, i.e. $P^\mu =p_c^\mu$. 
This gives\cite{bellini-blum}
\begin{equation}\label{B-general}
    \B = W\frac{\int d^4 r \int \frac{d^4k}{(2\pi)^4} \,D\left(r^\mu,k^\mu\right) S(r^\mu,k^\mu;P^\mu)}{\int d^4 r \,S(r^\mu,0;P^\mu)},
\end{equation}
where $D(r^\mu,k^\mu)$ is now the relativistic Wigner density for the bound state and
\begin{equation}
    W = \frac{(2\pi)^3m_c(2s_c+1)}{m_1m_2(2s_1+1)(2s_2+1)}.
\end{equation}
Notice that the denominator contains $S(r^\mu,0;p^\mu)$ without invoking the smoothness approximation. In fact, both of the event-mixed particles in the denominator and the coalesced nucleus are all assumed to have the same rest frame, i.e. $\q=0$ for this entire calculation and the on-shell approximation is exact (the mass of the coalesced nucleus is slightly less than the sum of the constituent masses, by the binding energy, but we neglect this here). We can therefore safely replace $P^\mu$ with $p^\mu$.

In the ETA \eqref{B-general} becomes
\begin{equation}\label{B-et}
    \B = W\frac{\int d^3 \rr \int \frac{d^3\kk}{(2\pi)^3} \,D\left(\rr,\kk\right) S(\rr,\kk;p^\mu)}{\int d^3 \rr \,S(\rr,0;p^\mu)},
\end{equation}
and if we were to apply the smoothness approximation, we would get the analog of the Koonin-Pratt formula for coalescence:
\begin{equation}
    \B \approx W \int d^3\rr \left|\phi(\rr)\right|^2 S(\rr;p^\mu).
\end{equation}

\section{The Smoothness Expansion} \label{sec3}
For the following analysis, we keep the dependence on the average CM momentum $p^\mu$ implicit, along with the sum over spin channels. Unless noted otherwise, all calculations are performed in the (approximate, via the on-shell approximation) PRF.

\subsection{Angle-Dependent Correlations} \label{sec3a}
The smoothness approximation is dependent on the source function varying slowly in $\q$. Schematically, every time we take a $\q$ derivative of the source function at $\q=0$ we will get some length scale $\e$ (for example in the blast wave model of Sec. \ref{sec5a} $\e$ is inverse temperature), and thus to investigate the corrections to the smoothness approximation we expand in small $q\e$:
\begin{equation}\label{s-expansion}
    S(\rr,\q) = S(\rr) + q_i S^i(\rr) + q_i q_j S^{ij}(\rr) + \oo{q^3\e^3}.
\end{equation}
This is normalized such that
\begin{equation}\label{snorm}
    \int d^3\rr\,S(\rr) = 1,
\end{equation}
and we also define
\begin{equation}
    \ei = \int d^3\rr\,S^i(\rr) , \qquad \eij = \int d^3\rr\,S^{ij}(\rr).
\end{equation}
Note that $S^i$ and $\ei$ will be order $\e$, while $S^{ij}$ and $\eij$ will be order $\e^2$.

Using the expansion \eqref{s-expansion}, we see that we can write the numerator of \eqref{cqp} as
\begin{multline}
    \int d^3 \rr \int \frac{d^3\kk}{(2\pi)^3} \,D_{\q}\left(\rr,\kk\right) S(\rr,\kk) \\ = \int d^3 \rr \left[ K(\rr,\q)\,S(\rr)+ K_i(\rr,\q)\, S^i(\rr) + K_{ij}(\rr,\q)\, S^{ij}(\rr) + \oo{q^3\e^3}\right]
\end{multline}
where we have the kernels
\begin{equation}\label{k}
    K(\rr,\q) \equiv \int \frac{d^3\kk}{(2\pi)^3} \,D_{\q}\left(\rr,\kk\right) = \left| \phi_{\q}(\rr) \right|^2,
\end{equation}
\begin{equation}\label{ki}
    K_i(\rr,\q) \equiv \int \frac{d^3\kk}{(2\pi)^3} \,k_i D_{\q}\left(\rr,\kk\right) = \frac{1}{2i}\left[\phi_{\q}^*(\rr)\,\partial_i \phi_{\q}(\rr) - \phi_{\q}(\rr)\,\partial_i \phi_{\q}^*(\rr)\right],
\end{equation}
and
\begin{multline}\label{kij}
    K_{ij}(\rr,\q) \equiv \int \frac{d^3\kk}{(2\pi)^3} \,k_ik_jD_{\q}\left(\rr,\kk\right) \\
    = \frac{1}{4}\left[\partial_i\phi_{\q}^*(\rr)\,\partial_j \phi_{\q}(\rr) + \partial_i\phi_{\q}(\rr)\,\partial_j \phi_{\q}^*(\rr) - \phi_{\q}^*(\rr) \partial_i \partial_j \phi_{\q}(\rr) - \phi_{\q}(\rr) \partial_i \partial_j \phi_{\q}^*(\rr) \right].
\end{multline}
Including the denominator into our calculation, noting that
\begin{equation}
    \int d^3 \rr \,S(\rr,\q) = 1 + q_i \ei + q_iq_j \eij + \oo{q^3\e^3},
\end{equation}
we get
\begin{multline}\label{gse}
    C(\q) = \int d^3 \rr \left[ K(\rr,\q)\,S(\rr)+ K_i(\rr,\q)\, S^i(\rr) + K_{ij}(\rr,\q)\,  S^{ij}(\rr)\right] \\
    - q_i \ei \int d^3 \rr \left[ K(\rr,\q)\,S(\rr)+ K_j(\rr,\q)\, S^j(\rr)\right] \\
    - q_iq_j (\eij-\ei\ej) \int d^3 \rr K(\rr,\q)\,S(\rr) + \oo{q^3\e^3}.
\end{multline}
This is the smoothness expansion; within the smoothness approximation we only retain
\begin{equation}
    \Cs(\q) = \int d^3\rr K(\rr,\q)S(\rr).
\end{equation}
The first three terms of Eq. \eqref{gse} are respectively the leading term, first-order correction, and second-order correction in the expansion of the numerator of \eqref{cqp}. The following two terms combine the numerator expansion with the first-order correction to the denominator of \eqref{cqp} up to second order overall. The final term is the leading term of the numerator multiplied by the second-order correction to the denominator, $-q_iq_j(\eij-\ei\ej)$.

In the expansion \eqref{gse}, the explicit factors of $q$ are the asymptotic momentum, but the kernels $K_i$ and $K_{ij}$ are moments of the Wigner function. For an interacting pair these moments are set by an effective momentum scale $q_{\rm eff}$ that includes the internal momentum of the pair weighted by the relevant moment of the source (i.e. $S$, $S_i$, or $S_{ij}$). At large $q$, this effective momentum scale will coincide with $q$, but it does not vanish as $q\to0$. In the kernel terms, the order $\oo{q^n\e^n}$ should therefore be read with $q$ referring to this effective scale.

As for the wavefunction with outgoing momentum $\q$, noting azimuthal symmetry around the axis of $\q$ we have
\begin{equation}\label{dwf}
    \phi_{\q}(\rr) = \sum_{\ell} i^{\ell} (2\ell + 1) R_{q,\ell}(r) P_{\ell}\left(\frac{\q\cdot\rr}{qr}\right),
\end{equation}
where $P_\ell$ are the Legendre polynomials and the functions $R_{q,\ell}$ are obtained by solving the Schrodinger scattering problem, subject to the boundary condition that $u_{q,\ell}(r)=rR_{q,\ell}(r)$ is regular at $r\to0$, and $u_{q,\ell}(r)\to e^{i\delta_\ell}\sin(qr-\ell\pi/2+\delta_\ell)/q$ as $r\to\infty$ (see e.g. \cite{higherpw,sakurai}). 
In practical computations, we have found the implementation in the Correlation Analysis Tool using the Schrodinger equation (CATS)\cite{cats} very useful\footnote{We note that the boundary condition at $r\to\infty$, recalled above in the standard form for the FSI problem, differs from the boundary condition implemented in CATS\cite{cats}, where the phase factor $e^{i\delta_\ell}$ was removed. For angle-dependent correlations this phase can be physically relevant, and we maintain it in the definition and in calculations.}. Since the wavefunction's directional dependence is contained entirely in $\q\cdot\rr$, we see that $K$, $K_i$, and $K_{ij}$ are respectively a polar scalar, vector, and tensor. Thus we get the relations $K(-\rr,-\q)=K(\rr,\q)$,  $K_i(-\rr,-\q)=-K_i (\rr,\q)$, and $K_{ij}(-\rr,-\q)=K_{ij}(\rr,\q)$.

In the case of identical particles, we must consider symmetrized and antisymmetrized wavefunctions, and sum over spin channels consistent with Pauli statistics. Specifically, for symmetrized wavefunctions only even $\ell$ contribute to \eqref{dwf}, and for antisymmetrized wavefunctions only odd $\ell$ contribute. We also multiply overall by a factor of $\sqrt{2}$ for normalization. Additionally, there are symmetry constraints on the source function for identical particles. Though wavefunction statistics are not relevant, the source function itself must be symmetric upon exchange of particles -- i.e. at the level of the source function, producing identical particles with separation $\rr$ and relative momentum $\q$ is the same as producing identical particles with separation $-\rr$ and relative momentum $-\q$. Because of this, $S^i(\rr)$ must be odd in $\rr$ since the overall term $q_i S^i(\rr)$ is odd in $\q$. Thus, for identical particles $\ei = 0$, and the only surviving first-order correction in the correlation function for identical particles is the $K_iS^i$ term (second-order terms involving $\ei$ also vanish, of course).

\subsection{Angle-Averaged Correlations} \label{sec3b}
We now consider the angle-averaged correlation
\begin{equation}\label{cqmag}
    C(q) = \frac{\idoq\int d^3 \rr \int \frac{d^3\kk}{(2\pi)^3} \,D_{\q}\left(\rr,\kk\right) S(\rr,\kk;p^\mu)}{\idoq\int d^3 \rr \,S(\rr,\q;p^\mu)}
\end{equation}
where $\oq$ is the solid angle in $\q$-space. This is often the reported quantity in femtoscopic studies, see e.g. \cite{alice-data,alice-strange1,alice-strange2,alice-strange3}. To study this, we look at the angle-averaging of terms appearing in the smoothness expansion. We will use the notation that functions which are denoted as a variation of $\bar{K}$ depend only on $r$ and $q$ (the magnitudes of the PRF three-vectors $\rr$ and $\q$ respectively), and functions which are denoted as a variation of $\bar{S}$ depend only on $r$.

First, we have
\begin{align}
    &\idoqa \, K(\rr,\q) = \kbar{0,0},\\
    &\idoqa\, q_i K(\rr,\q)=q\kbar{0,1}\rhat_i,\\
    &\idoqa\, q_iq_j K(\rr,\q) = \frac{q^2}{3}\kbar{0,0}\delta_{ij} + q^2 \kbar{0,2} Q_{ij}, \label{qiqjk}\\
    &\idoqa\, q_i K_j(\rr,\q) = q\kbar{1,0}\delta_{ij} + q \kbar{1,2} Q_{ij}, \label{qikj}\\
    &\idoqa\, K_{ij}(\rr,\q) = \kbar{2,0}\delta_{ij} + \kbar{2,2} Q_{ij}.
\end{align}
Here, $Q_{ij}=3\rhat_i\rhat_j-\delta_{ij}$ is the quadrupole tensor. The first index on $\bar{K}$ indicates which order of the expansion the kernel is from, and the second index denotes which degree moment it is attached to. The form of each of these integrals can be determined from symmetry considerations, and the coefficient on $\delta_{ij}$ in \eqref{qiqjk} can be determined from the spherical harmonic expansion of $K$. We note that while the LHS of \eqref{qikj} is not manifestly symmetric in the tensor indices, the only available non-symmetric term is of the form $\epsilon_{ijk}\rhat^k$ which is forbidden by evenness in $\rr$ (specifically, $q_i K_j(\rr,\q)$ must be even when $\rr,\q\to -\rr,-\q$ , so its angle average must be even in $\rr$). Finally, for a real spherically-symmetric FSI potential,
\begin{equation}\label{eq:aveKi}
    \idoqa\, K_i(\rr,\q) = 0.
\end{equation}
This is a special case of a general property: for real spherically-symmetric FSI potential, the angle average of any odd-order kernel $K_{i_1i_2...i_{2n+1}}$ (defined analogously to $K, K_i, K_{ij}$) vanishes. We refer to App. \ref{appC} for details.

Once the integrals above are performed, we define the following:
\begin{align}
    &\idor \, S(\rr) = \sbar{0,0} \\
    &\idor \,\delta_{ij} S^{ij}(\rr) = \sbar{2,0} \\
    &\idor \,Q_{ij} S^{ij}(\rr) = \sbar{2,2} \\
    &\idor \,Q_{ij}(\eij-\ei\ej) S(\rr) = \sbar{0,2} \\
    &\idor \, \ei \rhat_i S(\rr) = \sbar{0,1}\\
    &\idor \,Q_{ij}\varepsilon^i S^{j}(\rr) = \sbar{1,2} \\
    &\idor \,\delta_{ij}\varepsilon^i  S^{j}(\rr) = \sbar{1,0}.
\end{align}
Using all of this, we arrive at the smoothness expansion for $C(q)$:
\begin{equation}
    C(q) = \int dr\, r^2 \left[\left(1- \frac{q^2}{3}\tre \right)\ksbar{0,0} + \ksbar{2,0} + \ksbar{2,2}\right] + \oo{q^4\e^4}
\end{equation}
where $\tre=\delta_{ij}\varepsilon^{ij}$ (note the next nonzero term is fourth-order, as the third-order term vanishes under angle averaging as described above). We see here that the corrections to the correlation function begin at second-order.

We can compare $C(q)$ to the isotropic angle average of $C(\q)$, which has the smoothness expansion
\begin{multline}
    \idoqa C(\q) = \int dr\, r^2 \biggl[\left(1- \frac{q^2}{3}\tre + \frac{q^2}{3}\mage \right)\ksbar{0,0} + \ksbar{2,0} + \ksbar{2,2} \\
    -q\ksbar{0,1} - q\ksbar{1,0} - q\ksbar{1,2} - q^2 \ksbar{0,2}\biggr] + \oo{q^3\e^3}.
\end{multline}
where $\mage=\delta_{ij}\varepsilon^i\varepsilon^j$, and we have the anisotropic term
\begin{multline}\label{c-aniso}
    \Ca(q) = C(q) - \idoqa C(\q) \\
    = \int dr \, r^2 \left[-\frac{q^2}{3} \mage \ksbar{0,0} + q\ksbar{0,1} + q\ksbar{1,0} + q\ksbar{1,2} + q^2 \ksbar{0,2}\right] + \oo{q^3\e^3}.
\end{multline}
Eq.~(\ref{c-aniso}) is the difference between two observable quantities. This difference vanishes in the smoothness approximation, so it provides a method to assess the size of a subset of smoothness corrections in a model-independent way. These anisotropic corrections generally begin at first-order in the smoothness expansion; however for identical particles $\ei=0$, and thus the first order corrections vanish also here. In fact, the only surviving term of $\Ca(q)$ for identical particles is the $\ksbar{0,2}$ term.

\subsection{Free Particles} \label{sec3c}
The main application we have in mind for the analysis in this paper concerns realistic FSI; however, we first show the smoothness expansion in the context of free particles, in order to demonstrate the significance of each group of terms in the expansion.

\subsubsection{Distinguishable Particles} \label{sec3c1}
Free distinguishable particles should be uncorrelated, i.e. their correlation function should be identically 1 for all relative momenta. It is good to check that this holds to each order in the smoothness expansion; we have
\begin{equation}
    \phi_{\q}(\rr) = e^{i\q\cdot \rr},
\end{equation}
and thus
\begin{equation}
    K=1, \qquad K_i=q_i, \qquad K_{ij}=q_iq_j. 
\end{equation}
We then have the correlation function
\begin{equation}
    C (\q) = 1 + q_i\ei + q_iq_j\eij - q_i\ei - q_iq_j\eij =1=\Cs(\q).
\end{equation}
As expected, the contributions from the numerator and denominator of the correlation function perfectly cancel. The case with angle averaging is similar, and we get
\begin{equation}
    C(q) = \idoqa C(\q) = \Cs(q) = 1
\end{equation}
as required.

\subsubsection{Identical Particles} \label{sec3c2}
The situation is more interesting for identical particles. The respective symmetrized ($+$) and antisymmetrized ($-$) non-interacting wavefunctions are
\begin{equation}
    \phi^{\pm}_{\q,\mathrm{free}}(\rr) = \frac{1}{\sqrt{2}}\left(e^{i\q\cdot \rr} \pm e^{-i\q\cdot \rr}\right),
\end{equation}
which lead to the kernels $K=1\pm \cos(2\q\cdot \rr)$, $K_i=0$, and $K_{ij}=q_iq_j$. Additionally, we will always sum over spin channels. For identical particles with spin $s$, symmetric spin channels will have total weight $\frac{s+1}{2s+1}$ and antisymmetric spin channels will have total weight $\frac{s}{2s+1}$. Thus, enforcing the appropriate symmetrization conditions for bosons ($+$) and fermions ($-$), the overall kernels for free identical particles will be 
\begin{equation}
    K= 1\pm \frac{1}{2s+1} \cos(2\q\cdot \rr), \qquad K_i=0, \qquad K_{ij}=q_iq_j.
\end{equation}
We then have
\begin{equation}
     \Cs (\q) = 1 \pm \frac{1}{2s+1}\int d^3\rr\,\cos(2\q\cdot\rr)S(\rr)
\end{equation}
and the corrected correlation function
\begin{multline}
    C (\q) = 1 \pm \frac{1}{2s+1}\int d^3\rr\,\cos(2\q\cdot\rr)S(\rr) + q_iq_j\eij \\
    - q_iq_j\eij \left( 1\pm  \frac{1}{2s+1}\int d^3\rr\,\cos(2\q\cdot\rr)S(\rr) \right) + \oo{q^4\e^4}\\
    = 1 \pm \frac{1-q_iq_j\eij}{2s+1}\int d^3\rr\,\cos(2\q\cdot\rr)S(\rr) + \oo{q^4\e^4} \\
    =\Cs(\q) + q_iq_j\eij(1-\Cs(\q)) + \oo{q^4\e^4}.
\end{multline}
Thus, smoothness corrections do occur for non-interacting identical particles. We note that this expression is consistent with the exact\footnote{It is exact with respect to the smoothness \textit{and} equal-time approximations. On-shell corrections would still need to be considered.} formula for free identical particles
\begin{equation}\label{exact}
    C(\q) - 1 = (\Cs(\q)-1)\cdot \frac{\int d^3\rr\, S(\rr,0;p)}{\int d^3\rr\, S(\rr,\q;p)}
\end{equation}
which can be seen easily from, for example, \cite{pratt-smoothness} (that paper assumes that the two-particle source factorizes, but this is not necessary for the above formula. Additionally, that paper works only with spin-0 bosons, but it can be generalized easily to fermions and higher spins with the proper $\frac{1}{2s+1}$ factor). Considering the symmetry properties of the source function for identical particles discussed in Sec. \ref{sec3a}, it is clear from Eq. \eqref{exact} that all odd-order corrections vanish for free identical particles, even at the level of angle-dependent correlations.

With angle averaging, we note the integrals
\begin{align}
    &\idoqa \cos(2\q\cdot \rr) = j_0(2qr),\\
    &\idoqa q_iq_j \cos(2\q\cdot \rr) = \frac{q^2}{3}j_0(2qr)\delta_{ij} - \frac{q^2}{3}j_2(2qr)Q_{ij},
\end{align}
where $j_n$ are the spherical Bessel functions, so
\begin{equation}
    \Cs(q) = 1\pm \frac{1}{2s+1}\int dr\, r^2 \, j_0(2qr)\sbar{0,0}
\end{equation}
and
\begin{multline}
    C (q) = 1\pm \frac{1}{2s+1}\int dr\, r^2 \, j_0(2qr)\sbar{0,0} +  \frac{q^2}{3}\tre \\
    - \frac{q^2}{3}\tre \left( 1 \pm \frac{1}{2s+1}\int dr\, r^2 \, j_0(2qr)\sbar{0,0}\right) + \oo{q^4\e^4}\\
    = 1 \pm \frac{1}{2s+1}\left(1-\frac{q^2}{3}\tre\right)\int dr\, r^2 \, j_0(2qr)\sbar{0,0} + \oo{q^4\e^4} \\
    =\Cs(q) + \frac{q^2}{3}\tre(1-\Cs(q)) + \oo{q^4\e^4}.
\end{multline}
Finally, the anisotropic correction term is also nonzero for free identical particles:
\begin{equation}
    \Ca(q) = \mp \frac{q^2}{3(2s+1)}\int dr\, r^2\, j_2(2qr)\sbar{0,2} + \oo{q^4\e^4}.
\end{equation}

\subsection{Coalescence} \label{sec3d}
Recalling the form of the coalescence factor in \eqref{B-et}, the smoothness expansion for coalescence only has the numerator contributions,
\begin{equation}
    \B = W \int d^3 \rr \left[ K(\rr)\,S(\rr)+ K_i(\rr)\, S^i(\rr) + K_{ij}(\rr)\, S^{ij}(\rr) + \oo{q^3\e^3}\right].
\end{equation}

We use a Gaussian wavefunction
\begin{equation}\label{gauss-wf}
    \phi(\rr) = \left(\frac{1}{\pi b^2}\right)^{3/4} e^{-\frac{r^2}{2b^2}}
\end{equation}
to investigate qualitative behavior. The kernels are then conveniently
\begin{equation}
    K = |\phi|^2 = \left(\frac{1}{\pi b^2}\right)^{3/2} e^{-\frac{r^2}{b^2}}, \qquad K_i = 0, \qquad K_{ij} = \frac{|\phi|^2}{2b^2}\delta_{ij} = \frac{1}{2b^2}\left(\frac{1}{\pi b^2}\right)^{3/2} e^{-\frac{r^2}{b^2}}\delta_{ij}.
\end{equation}
Note that odd-order kernels will be zero for real wavefunctions (or real up to an overall phase) as the Wigner density (Eq. \eqref{dq}) will be even in $\kk$.

Since we have chosen our wavefunction to be rotationally symmetric, $|\phi|^2$ can be pulled out of the $\Omega_{\rr}$ integral, and we get
\begin{equation}
    \B = W \int dr\, r^2\, |\phi|^2 \left(\sbar{0,0} + \frac{1}{2b^2} \sbar{2,0} + \oo{q^4\e^4}\right).
\end{equation}
The smoothness corrections are greater for a narrower wavefunction. For wavefunctions with nontrivial angular dependence other $\bar{S}$ terms can contribute, however for a real central interaction the ground state wavefunction will have spherical symmetry and can be chosen to be real. Thus in this case the smoothness correction will only involve $\sbar{2,0}$ until fourth order, though the kernel $\kbar{2,0}$ will not in general be proportional to $|\phi|^2$ as it is in the Gaussian case.

\section{The On-Shell Expansion} \label{sec4}
We now look at the corrections to the on-shell approximation. These corrections are purely kinematical; additionally, recalling that the on-shell approximation is exact for coalescence, we need only look at corrections to the correlation function. For the kinematic calculations in the first part of this section, primed quantities are in the PRF frame, and unprimed quantities are in the lab frame; additionally, unbolded momenta will refer to four-vectors as opposed to the magnitude of a three-vector.

Recall the kinematic definitions \eqref{mom-decomp} and \eqref{p-vs-P}, which introduce the true average CM momentum $P^\mu =\frac{p_1^\mu+p_2^\mu}{2}\equiv (P^0,\p)$ and the on-shell average pseudo-CM momentum $p^\mu =(p^0, \p)$ with $(p^0)^2 = m^2 + \p^2$ where we write $m\equiv \frac{m_1+m_2}{2}$ as a shorthand. Some algebra gives
\begin{equation} \label{P0p0}
    P^0 = p^0 + \frac{\kappa \q^2}{2p^0} + \oo{q^3/m^2}
\end{equation}
where
\begin{equation}
    \kappa = \frac{m^2}{m_1m_2}.
\end{equation}
In the equal-mass case, $\kappa=1$. Note that here we bold $\q^2$ to avoid confusion with $q^\mu q_\mu=-\q^2$.

In the on-shell approximation, we analyze in a pseudo-PRF (we denote this frame with a double prime) such that $p''^\mu=\left(m, \boldsymbol{0}\right)$, while without this approximation we would be working in the true PRF where $P'^\mu=\left(\sqrt{(P^0)^2 - \p^2},\boldsymbol{0}\right)$.  The transformation from the pseudo-PRF to the true PRF is a boost with $\beta_i = -\frac{\kappa \q^2}{2m^2p^0} p_i + \oo{q^3/m^3}$ which begins at order $q^2$, and thus the Lorentz factor is unity up until order $q^4$. So, the Lorentz transformation from the pseudo-PRF to the true PRF is
\begin{equation}
    v'^\mu =  \begin{pmatrix}
1 & \frac{\kappa \q^2}{2m^2p^0}\p \\
\frac{\kappa \q^2}{2m^2p^0}\p & I \\
\end{pmatrix} v''^\mu + \oo{q^3/m^3}.
\end{equation}
A corollary of this is that $\q''=\q'$ (and in the case of $\kk$ in the Wigner function integral $\kk''=\kk'$) up to third order in relative momentum. This means that to correct the on-shell approximation to second order (leading order) we need only replace $p^\mu$ with $P^\mu$, replace $\alpha_0$ with $\alpha$ (recall the definitions \eqref{alpha} and \eqref{alpha0}), and consider the transformation of $r^\mu$ in the source function (remembering that this approximation is made only in the calculation of the source function -- the Wigner function is already by assumption expressed in the true PRF). For replacing $\alpha_0$ with $\alpha$, we have
\begin{equation}
    \alpha = \alpha_0 - \frac{\alpha_0\q^2}{m_1m_2} + \oo{q^3/m^3}.
\end{equation}
This will create corrections to the source function as the two-particle source will be sensitive to $\alpha$, which describes how momentum is divided between the particles. This is irrelevant for equal mass particles, where $\alpha = 0$ for all $q$.

Considering the above, we also have that the first on-shell corrections will be of order $q^2$; specifically, some will be order $q^2/m^2$, while part of the correction for replacing $P^\mu$ with $p^\mu$ will be order $q^2\e/m$. This also means that to order $q^2$ all of the on-shell corrections are corrections to the smooth correlation function, as all combinations between on-shell and smoothness corrections will be of order $q^3$ or higher. Now,
\begin{multline}\label{sprime}
    S(\rr', \q';P^\mu) = \int dr'^0 S(r'^\mu,q'^\mu;P^\mu)\mid_{q'^0=0} \\
    = \int dr''^0 S(r''^\mu,q''^\mu;p^\mu)\mid_{q''^0=0} - \frac{\kappa \q^2}{2m^2p^0} (\p\cdot \rr'') \int dr''^0 \partial_{r''^0} S(r''^\mu,q''^\mu;p^\mu)\mid_{q''^0=0} \\ - \frac{\kappa \q^2}{2m^2p^0} p^i \int dr''^0 r''^0 \partial_{r''^i} S(r''^\mu,q''^\mu;p^\mu)\mid_{q''^0=0} + \frac{\kappa \q^2}{2p^0} \int dr''^0 \partial_{p^0} S(r''^\mu,q''^\mu;p^\mu)\mid_{q''^0=0} \\
    - \frac{\alpha_0 \q^2}{m_1m_2}\int dr''^0 \partial_{\alpha} S(r''^\mu,q''^\mu;p^\mu)\mid_{q''^0=0} + \oo{q^3\e^2/m} \\
    \equiv S(\rr'',\q'';p^\mu) + \q^2 S'(\rr'',\q'';p^\mu)+ \oo{q^3\e^2/m}.
\end{multline}
Note that the first correction term is zero since it is a boundary term, and as mentioned the $\alpha_0$ term will be zero for particles of equal masses. Since we are not concerned with combining smoothness and on-shell corrections, what will be useful to us is
\begin{multline}
    S(\rr',0;P^\mu) = S(\rr'',0;p^\mu) + \q^2 \cdot S'(\rr'',0;p^\mu)+ \oo{q^3\e^2/m} \\
    \equiv S(\rr'';p^\mu) + \q^2 S'(\rr'';p^\mu)+ \oo{q^3\e^2/m} .
\end{multline}

We are now done working with frame transformations, so we return to working entirely in the pseudo-PRF and suppressing the $p^\mu$ argument of $S$ and $S'$ ($p^0$ and $\p$ will of course still refer to the components of $p^\mu$ in the lab frame). We will also return to using $q^2 = \q^2$ as before. We ignore smoothness corrections as they do not affect this part of the calculation. The correlation function is
\begin{multline} \label{os-expansion}
    C(\q) = \frac{\int d^3 \rr \, K(\rr,\q)\left(S(\rr) + q^2 S'(\rr)\right)}{\int d^3 \rr \left(S(\rr) + q^2 S'(\rr)\right)} + \oo{q^3\e^2/m} \\
    = \left(1-q^2 \varepsilon'\right) \int d^3 \rr  \, K(\rr,\q)S(\rr) + q^2\int d^3\rr\,  K(\rr,\q)S'(\rr) + \oo{q^3\e^2/m} \\
    = \left(1-q^2\varepsilon'\right) C_\mathrm{on-shell}(\q) + q^2\int d^3\rr  \, K(\rr,\q) S'(\rr) + \oo{q^3\e^2/m},
\end{multline}
where
\begin{equation}
    \varepsilon' = \int d^3 \rr\, S'(\rr).
\end{equation}
Note that only the $\partial_{p^0}$ and $\partial_\alpha$ terms of \eqref{sprime} contribute to $\varepsilon'$, as the $\partial_{r''^i}$ term integrates to vanishing boundary terms in the direction of $\p$.

The angle-averaged correlation is simple,
\begin{equation}
    C(q) = \int dr\, r^2 \, \kbar{0,0} \left[(1-q^2 \varepsilon')\sbar{0,0} +  q^2 \sbar{}' \right] + \oo{q^3\e^2/m}
\end{equation}
where
\begin{equation}
    \sbar{}' = \idor S'(\rr).
\end{equation}
The on-shell corrections only have $q$-dependence through $q^2$, so they are isotropic in $\q$ and generate no contribution to $C(q) - \idoqa C(\q)$.

For free distinguishable particles, analogous calculations to those done in Sec. \ref{sec3c} show that all corrections vanish as we would expect. For identical particles, we get
\begin{equation}
    C(\q) = 1 \pm \frac{1}{2s+1}\int d^3\rr \, \cos(2\q\cdot\rr)\left[(1-q^2 \varepsilon')S(\rr)+q^2 S'(\rr)\right] + \oo{q^3\e^2/m}
\end{equation}
and
\begin{equation}
    C(q) = 1  \pm \frac{1}{2s+1}\int dr\, r^2 j_0(2qr)\left[(1-q^2\varepsilon')\sbar{0,0}+q^2\sbar{}'\right] + \oo{q^3\e^2/m}.
\end{equation}

\section{Results for the Blast Wave Model} \label{sec5}
We show example calculations of the smoothness corrections using the blast wave model for the source function, considering angle-averaged pp correlations with realistic FSI, and deuteron coalescence with a Gaussian deuteron wavefunction. We focus on pp correlations both because they are dominated by FSI effects (the main use case of our formalism) and also for their applicability in studying the coalescence-correlation framework. 

\subsection{The Blast Wave Model} \label{sec5a}
The blast wave model\cite{bwm} is a phenomenological relativistic hydrodynamic model for a single-particle source function $S_1(x^\mu,p^\mu)$, corresponding to the production of a particle at spacetime coordinate $x^\mu$ and four-momentum $p^\mu$. Since it is a single-particle source function, for the pair source we assume that the two particles are produced independently of each other (this is equivalent to density matrix factorization). Then we have, in the case with two different particles $a$ and $b$,
\begin{multline}\label{s-bwm}
    S(\rr, \q;p^\mu) = \mathcal{N}\int dr^0 S(r^\mu,q^\mu;p^\mu)\mid_{q^0=0} \\
    = \mathcal{N}\int dr^0 \int d^4x S_{1,a}\left( x^\mu+\frac{r^\mu}{2},(1+\alpha) p^\mu+\q\right)S_{1,b}\left(x^\mu-\frac{r^\mu}{2},(1-\alpha)p^\mu-\q\right),
\end{multline}
where $\mathcal{N}$ is a normalization constant to satisfy the condition \eqref{snorm}:
\begin{equation}
    \frac{1}{\mathcal{N}} = \int d^4r \int d^4x S_{1,a}\left( x^\mu+\frac{r^\mu}{2},(1+\alpha) p^\mu\right)S_{1,b}\left(x^\mu-\frac{r^\mu}{2},(1-\alpha)p^\mu\right).
\end{equation}

We define the single-particle source function as a function of the lab-frame coordinates, which must be transformed to the PRF for use in \eqref{s-bwm}. Letting $z$ be the axis of the beam, we use coordinates $\tau = \sqrt{t^2-z^2}$, $\rho = \sqrt{x^2+y^2}$, $\eta = \mathrm{arctanh}(z/t)$, and the azimuthal angle $\varphi$. Then\cite{bwm,bellini-blum}
\begin{align}
    S_1^{\mathrm{lab}}(x^\mu,p^\mu)& = \left(p^0\cosh(\eta)-p^z\sinh(\eta)\right) J(\tau) \Theta(R_0 - \rho) e^{-p^\mu u_\mu / T}, \label{bwm-slab} \\
u^\mu(x^\mu) & =  (\cosh\eta \cosh\eta_t,\, \sinh\eta_t \cos\varphi,\, \sinh\eta_t \sin\varphi,\, \sinh\eta \cosh\eta_t), \label{u} \\
\eta_t(\rho) & = \mathrm{arctanh}\left( \frac{\rho^n}{R_0^n} \beta_S \right), \\
J(\tau) & = \frac{1}{\sqrt{2\pi}\,\Delta\tau}
e^{-\frac{(\tau - \tau_0)^2}{2\,\Delta\tau^2}},
\end{align}
where $\Theta$ denotes the Heaviside step function. This model has 6 parameters: $\tau_0$,  $\Delta\tau$, $R_0$, $\beta_S$, $n$, and the temperature $T$.  The only difference in the source function for different particle species is the mass which appears in the momentum $p^\mu$ and the $\alpha$ factor. Note that we write the Cooper-Frye (CF) factor\cite{cooper-frye} here as $p^0\cosh(\eta)-p^z\sinh(\eta)$ which can be applied off-shell\cite{current}. On-shell, where $p^\mu$ is parameterized as $p^\mu = (m_t \cosh Y, p_t \cos\Phi, p_t\sinh\Phi, m_t \sinh Y)$ with $m_t = \sqrt{m^2 + p_t^2}$, this factor is more commonly seen as $\frac{m_t}{m}\cosh(\eta-Y)$.

For simplicity, we restrict ourselves to identical particles. 
We choose a coordinate system such that $\Phi=0$ (which we can do WLOG) and specialize to midrapidity ($Y=0$), giving us $p^\mu =(m_t,p_t,0,0)$. We refer to App. \ref{appB} for details on the smoothness and on-shell corrections for this model.

\subsection{Correlations} \label{sec5b}
In the following calculations, we used CATS\cite{cats} for numerical wavefunctions (adding in the appropriate complex phase shifts, as mentioned in Sec. \ref{sec3a}), using the Argonne v18 potential\cite{av18} to model FSI. See App. \ref{appA} for details on the calculation of angle-averaged kernels. We refer to \cite{alice-bwm1,alice-bwm2,alice-coal-bwm} for typical blast wave parameter fits to ALICE collisions (which generally focus on PbPb collisions). We also fit the data in Figure 1 of \cite{alice-data} for pp collisions with $p_t\approx0.9$ GeV to the blast wave model with parameters $R_0=\tau_0=2.0$ fm, $\Delta\tau=1.5$ fm, $\beta_S=0.5$, $n=2$, and $T=150$ MeV. 

\begin{figure}
    \centering
    \includegraphics[width=0.7\linewidth]{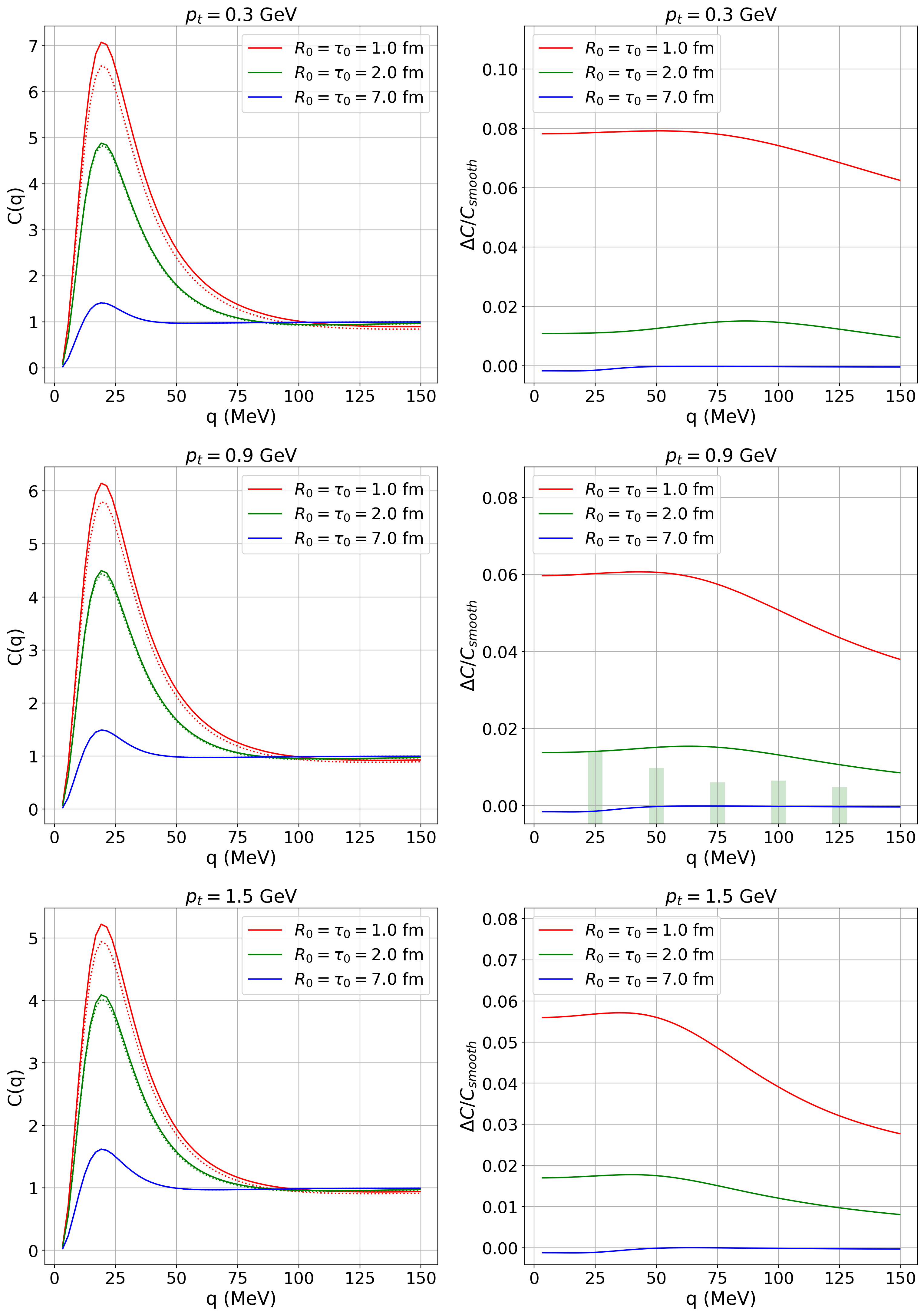}
    \caption{Left: $C(q)$ (solid) and $\Cs(q)$ (dotted) for various source sizes and transverse momenta. Right: $(C(q)-\Cs(q))/\Cs(q)$ for the corresponding plots on the left. The green curves in the central row are based on explicit fits to data in \cite{alice-data}\footnote{The plot presented here does not exactly match the plot in \cite{alice-data}. This is because we include only the pure p-p correlation function, while \cite{alice-data} includes p-$\Lambda$ pairs for which the hyperon decays into a proton prior to reaching the detector.}, and the error bars on the central right panel correspond to the modeling uncertainty reported there. Fixed parameters are $\Delta\tau=1.5$ fm, $\beta_S=0.5$, $n=2$, and $T=150$ MeV.}
    \label{fig:corr-size}
\end{figure}

\begin{figure}
    \centering
    \includegraphics[width=1\linewidth]{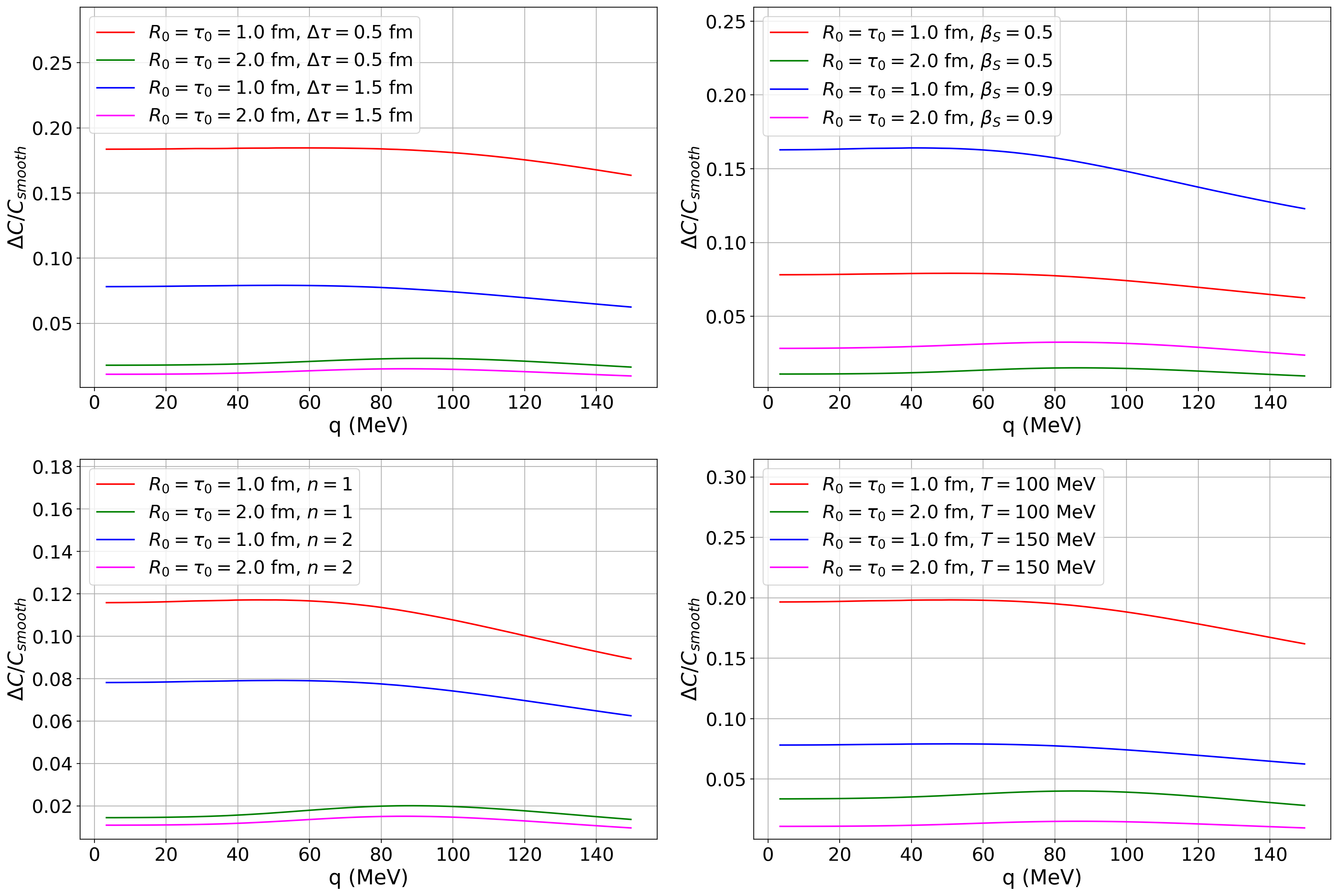}
    \caption{$(C(q)-\Cs(q))/\Cs(q)$ for various blast wave parameters and transverse momentum $p_t=0.3$ GeV. Unless indicated otherwise, default parameters are $\Delta\tau=1.5$ fm, $\beta_S=0.5$, $n=2$, $T=150$ MeV.}
    \label{fig:corr-other}
\end{figure}

We plot the correlation functions and corrections for various source sizes and transverse momenta in Fig. \ref{fig:corr-size}, using the $\Delta\tau$, $\beta_S$, $n$, and $T$ values from the previously mentioned fit to the data in \cite{alice-data}. We use $R_0=\tau_0=2.0$ fm as an approximate picture of pp collisions, $R_0=\tau_0=7.0$ fm for PbPb collisions, and $R_0=\tau_0=1.0$ fm to demonstrate the effects of a smaller source. The green curve in the center row is the specific fit for the data. We found the smoothness corrections for pp source parameters to be slightly larger than the reported uncertainty of the fit in \cite{alice-data}, of order 0.5\% (see the error bars in Fig. \ref{fig:corr-size}). For smaller sources, the corrections are several percent. Fig. \ref{fig:corr-other} shows the effects of varying the other blast wave parameters for different source sizes at fixed $p_t=0.3$ GeV. Additionally, we found the anisotropic corrections \eqref{c-aniso} to be universally small -- sub-percent even for small sources, and sub-permil for pp source parameters.


The size of the source $R_0\approx\tau_0$ is a primary factor for the magnitude of both of the correlation function itself, as well as the relative corrections to it due to the smoothness expansion. For smaller sources, the correlation function and its relative corrections become larger. Specifically, the corrections become large when the size of the source becomes comparable to the characteristic length of the interaction. This scaling is analyzed further at the level of the source function in App. \ref{appB}. 
We also found that the smoothness corrections were larger at smaller transverse momenta. 

For small sources, a smaller $\Delta\tau$ generated larger corrections, as it made the source more localized at a small $r$. Lowering $n$ and increasing $\beta_S$ also weakened the smoothness approximation, as it made the flow velocity $u^\mu$ more extreme. The temperature also plays a direct role in the size of the corrections, as the thermal contribution scales as $q^2/T^2$, and thus smaller temperatures 
bring larger corrections.

As mentioned previously, terms in the smoothness expansion are controlled by some effective momentum scale $q_{\rm eff}$ which tends to the asymptotic momentum at large $q$. For the pp-like source, at small $q$ the effective momentum scale is tens of MeV (as can be seen from the size of the corrections). Higher terms in the expansion will carry at least one additional factor of $(q_{\rm eff}/T)^2$ (with some factorial prefactor as well, coming from the Taylor expansion of the exponential). Therefore, for $q\ll T$ we expect the corrections presented to capture the full order-of-magnitude smoothness correction. As $q$ (and thus $q_{\rm eff}$) approaches $T$, in our calculations either 100 or 150 MeV, higher-order corrections could be of the same order of magnitude as the corrections presented here.

Overall, we found the smoothness corrections to be dominant compared to the on-shell corrections, and specifically the thermal factor was the main contributor to the smoothness corrections. The on-shell corrections and the CF smoothness corrections were smaller; this can be explained by the thermal correction being of the form $q^2/T^2$ while the remaining corrections are of the forms $q^2/(mT)$ or $q^2/m^2$. 
For lower-$m$ particles, like pions, this balance is likely to change. An analysis of free identical boson correlations, using the pion mass and a blast-wave source with pp-like parameters, shows that on-shell corrections dominate the overall corrections, generating corrections of 1-2\% at mid-range $p_t$ between 0.3 and 0.9 GeV. The anisotropic correction is slightly smaller, at the level of about 0.5\%, but is in principle observable.

\subsection{Coalescence} \label{sec5c}

\begin{figure}
    \centering
    \includegraphics[width=1\linewidth]{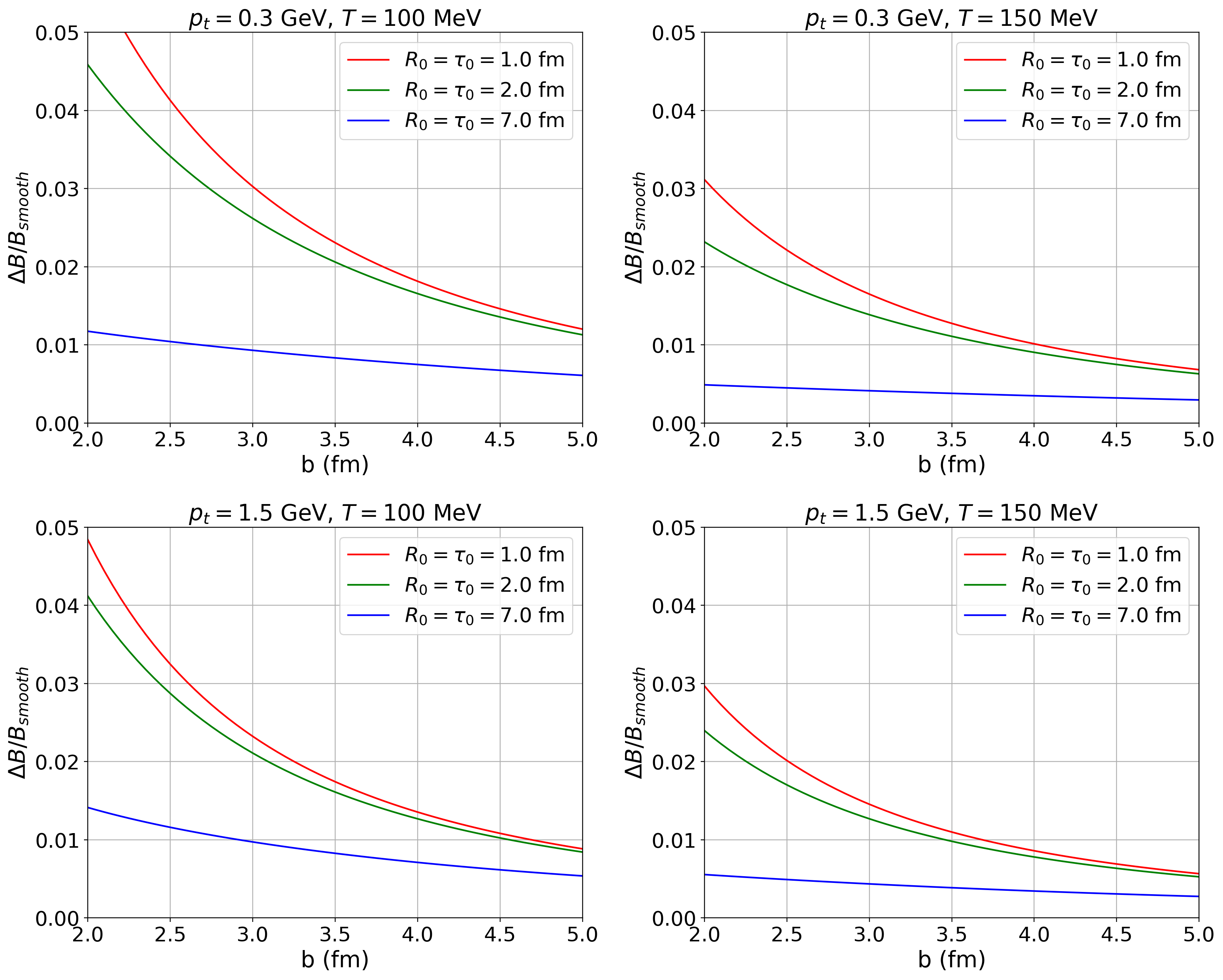}
    \caption{$(\B-\B_\mathrm{smooth})/\B_\mathrm{smooth}$ versus $b$ for coalescence with a Gaussian wavefunction of width $b$. The ratio is plotted for various source sizes, transverse momenta, and temperatures. Fixed parameters are $\Delta\tau=1.5$ fm, $\beta_S=0.5$, and $n=2$. Note that the plots on the right column correspond to the parameters featured in the top and bottom rows of Fig. \ref{fig:corr-size}.}
    \label{fig:coal-size}
\end{figure}

We now turn to the relative correction to the coalescence factor $(\B-\B_\mathrm{smooth})/\B_\mathrm{smooth}$ for a Gaussian wavefunction \eqref{gauss-wf} of width $b$. These are plotted for various $b$, source sizes, and temperatures in Fig. \ref{fig:coal-size}. For the deuteron, $b\approx 3.5$ fm\cite{bellini-blum}, which is centered in these plots. As expected, corrections are greater for smaller source sizes, smaller $b$, and smaller $T$. For the deuteron, the corrections are of order percent, below typical measurement uncertainties\cite{alice-coal-bwm}. Higher-order terms will come with additional factors of $1/(bT)^2$ which approaches unity around $b=2$ fm for $T=100$ MeV and around $b=1.3$ fm for $T=150$ MeV. For larger $b$, the second-order correction is dominant.

\section{Discussion} \label{sec6}
We have derived corrections to the smoothness and on-shell approximations for both correlation and coalescence calculations. These corrections can be used to test the validity of the smoothness approximation for a given source model and FSI. Additionally, we have presented a combination of observables \eqref{c-aniso} which can be used to empirically test for the presence of a particular subset of smoothness corrections.

In the specific case of the blast wave model with parameters relevant for high-energy pp collisions, the corrections to the angle-averaged correlation can reach about 2\% (see green curve in the right panels of Fig.~\ref{fig:corr-size}), above the nominal experimental and modeling uncertainty for proton-proton correlations reported in \cite{alice-data} (Fig.1(a) {\it there}). 
For parameters relevant to PbPb collisions, the corrections are smaller. Other, potentially more realistic source models can predict similar correlation functions\cite{book}, but could in principle exhibit different smoothness and on-shell corrections which can be tested using our method. Because of this model-dependence, one should be especially careful when interpreting these corrections; applications of these corrections within a universal-source analysis may still be subject to the source-universality issues raised in  \cite{epelbaum}.

Other commonly used approximations have also been shown to induce effects at the percent level, such as the Lednicky-Lyuboshitz asymptotic approximation\cite{romanenko-bellini} and the use of a Gaussian source function\cite{levy-stable}. Although, for the particular model considered here, these corrections are larger than the smoothness and on-shell corrections, we emphasize that those analyses themselves implicitly rely on the smoothness and on-shell approximations. The corrections discussed in the present work should therefore be considered alongside those associated with such approximations in order to obtain a complete assessment of systematic effects. 

We showed that 
in the case of angle-averaged correlations, the leading smoothness corrections are at order $q^2$, but for angle-dependent correlations one must also consider order $q$ terms. Even for identical particles, the first-order correction could be an order of magnitude larger than the corrections we saw in the angle-averaged correlation, in the range of a few percent for pp-like blast wave sources. For distinguishable particles, a first-order term also appears in the anisotropic correction \eqref{c-aniso}; this is one of several terms which are present in the smoothness expansion for distinguishable particles but not for identical particles, and may be significant for, e.g., p$\Lambda$ correlations.

In some cases, there were also large second-order corrections individually in the out, side, and long axes, but largely canceled out under angle averaging. These terms do not always show up in  \eqref{c-aniso} as they affect higher moments of the correlation function. It does not, however, appear straightforward to isolate these terms in an observable similar to \eqref{c-aniso}.

Finally, in our calculations we have used the equal-time approximation, which must also be verified in a given model/interaction in order to safely use the Koonin-Pratt formalism. An initial inspection of this approximation indicates that corrections to the ETA may be close to or larger than the smoothness corrections for the model we have considered. Leading corrections should scale as $q^2t/m$ where $t$ is some characteristic time scale of the source (see, for example, \cite{finite-size}). In the blast wave model, $t\sim \tau_0$ which is around 2 fm for pp sources. Thus, equal-time corrections would be of order $10^{-5}\, \mathrm{MeV}^{-2}\, q^2$ which is in the range of several percent when $q\sim 50$ MeV. We hope to investigate this approximation in future work.

\section*{Acknowledgments}
We thank D. Mihaylov and L. Fabbietti for helpful discussions. This work was supported by the Minerva foundation with funding from the Federal German Ministry for Education and Research. 

\section*{Data Availability}
The data and code that support the findings of this article are openly available \cite{zenodo-repo}.

\medskip

\bibliographystyle{apsrev4-2}
\bibliography{refs}

\appendix
\section{Technical Details on Calculating Averaged Kernels} \label{appA}
We have three objects of the form
\begin{equation}
    I_{ij} = A\delta_{ij} + B Q_{ij}.
\end{equation}
Since there are only two independent terms, it is sufficient to calculate just two elements of $I_{ij}$. Using a cartesian coordinate system where $\rr$ is in the $\zhat$ direction, we have that
\begin{equation}
    I_{zz} = A + 2B, \qquad I_{xx} = A - B
\end{equation}
so
\begin{equation}
    A = \frac{1}{3}\left(I_{zz} + 2I_{xx}\right), \qquad B = \frac{1}{3}\left(I_{zz} - I_{xx}\right).
\end{equation}

We consider a wavefunction $\phi_{\q}$ which is azimuthally symmetric (in the case of a central interaction, one can choose an axis such that this is the case). Let $\theta$ be the angle between $\rr$ and $\q$, let $\varphi$ be the azimuthal angle, and let $\mu=\cos(\theta)$. Then we have (dropping the subscript for simplicity):
\begin{equation}
    \q\q \supset q^2\mu^2\zhat\zhat + q^2(1-\mu^2)\cos^2(\varphi)\xhat\xhat,
\end{equation}
\begin{equation}
    \q\nabla\phi \supset q\mu \partial_r \phi \zhat\zhat + q(1-\mu^2) \cos^2(\varphi)\frac{\partial_\mu \phi}{r}\xhat\xhat,
\end{equation}
\begin{equation}
    \nabla\phi^*\nabla\phi \supset |\partial_r \phi|^2 \zhat\zhat + (1-\mu^2)\cos^2(\varphi)\frac{|\partial_\mu \phi|^2}{r^2}\xhat\xhat,
\end{equation}
and
\begin{equation}
    \nabla\nabla\phi \supset \partial_r^2\phi \zhat\zhat + \frac{(1-\mu^2) \cos^2(\varphi)\partial_\mu^2 \phi + r \partial_r \phi - \mu \partial_\mu \phi}{r^2} \xhat\xhat.
\end{equation}
Then,
\begin{equation}
    q_zq_zK = q^2 \mu^2 |\phi|^2,\qquad q_x q_x K = q^2(1-\mu^2)\cos^2(\varphi) |\phi|^2,
\end{equation}
\begin{equation}
    q_z K_z = \Im \left[ q\mu \phi^* \partial_r \phi\right], \qquad q_x K_x = \Im \left[q(1-\mu^2) \cos^2(\varphi)\frac{\phi^* \partial_\mu \phi}{r} \right],
\end{equation}
and
\begin{multline}
   K_{zz} = \frac{1}{2}|\partial_r \phi|^2 - \frac{1}{2}\Re \left[ \phi^* \partial_r^2\phi \right], \\
   K_{xx} = \frac{(1-\mu^2)}{2r^2}\cos^2(\varphi)|\partial_\mu \phi|^2 - \frac{1}{2r^2}\Re \left[ (1-\mu^2) \cos^2(\varphi)\phi^* \partial_\mu^2 \phi + r \phi^* \partial_r \phi - \phi^* \mu \partial_\mu \phi \right].
\end{multline}

Integrating over $\varphi$ is simple, noting that the $\varphi$-average of $\cos^2(\varphi)$ is $\frac{1}{2}$, and we arrive at the expressions:
\begin{equation}
    \kbar{0,0} = \frac{1}{2}\int_{-1}^1 d\mu\, |\phi|^2,
\end{equation}
\begin{equation}
    \kbar{0,2} = \frac{1}{12}\int_{-1}^1 d\mu \, (3\mu^2-1) |\phi|^2 = \frac{1}{6}\int_{-1}^1 d\mu \, P_2(\mu)|\phi|^2 ,
\end{equation}
\begin{equation}
    \kbar{1,0} = \frac{1}{6}\int_{-1}^1 d\mu \left( \mu \Im\left[\phi^* \partial_r \phi\right] + \frac{1-\mu^2}{r} \Im\left[\phi^* \partial_\mu \phi\right] \right),
\end{equation}
\begin{equation}
    \kbar{1,2} = \frac{1}{6}\int_{-1}^1 d\mu \left( \mu \Im\left[\phi^* \partial_r \phi\right] - \frac{1-\mu^2}{2r} \Im\left[\phi^* \partial_\mu \phi\right] \right),
\end{equation}
\begin{equation}
    \kbar{2,0} = \frac{1}{12}\int_{-1}^1 d\mu \Bigg( |\partial_r \phi|^2 - \Re\left[\phi^* \partial_r^2\phi\right] + \frac{2}{r^2}\Big( (1-\mu^2) |\partial_\mu \phi|^2 - \Re\left[r\phi^* \partial_r \phi\right] \Big)\Bigg),
\end{equation}
and
\begin{equation}
    \kbar{2,2} = \frac{1}{12}\int_{-1}^1 d\mu \Bigg( |\partial_r \phi|^2 - \Re\left[\phi^* \partial_r^2\phi\right] - \frac{1}{r^2}\Big( (1-\mu^2) |\partial_\mu \phi|^2 - \Re\left[r\phi^* \partial_r \phi\right] \Big)\Bigg).
\end{equation}
For $\kbar{2,0}$ and $\kbar{2,2}$ we used integration by parts to note that
\begin{multline}
    \idoqa K_{xx} = \frac{1}{4r^2}\int_{-1}^{1} d\mu \left(\frac{(1-\mu^2)}{2}|\partial_\mu \phi|^2 - \Re \left[ \frac{(1-\mu^2)}{2} \phi^* \partial_\mu^2 \phi + r \phi^* \partial_r \phi - \mu \phi^* \partial_\mu \phi \right] \right) \\
    = \frac{1}{4r^2} \int_{-1}^{1} d\mu \Big( (1-\mu^2)|\partial_\mu \phi|^2 - \Re \left[r \phi^* \partial_r \phi\right] \Big).
\end{multline}

Finally, our one remaining kernel is simply
\begin{equation}
    \kbar{0,1} = \idoqa \frac{q_z}{q} K = \frac{1}{2} \int_{-1}^{1} d\mu\,\mu |\phi|^2 = \frac{1}{2}\int_{-1}^1 d\mu \, P_1(\mu)|\phi|^2.
\end{equation}

\section{Smoothness and On-Shell Corrections to the Blast Wave Model}\label{appB}
\subsection{Smoothness Corrections}
Recall that we take $p^\mu =(m_t,p_t,0,0)$. Using the usual out-side-long terminology, we write $\q=(q_o,q_s,q_l)$ in the PRF, which gives $q^{\mu,\mathrm{lab}}=\left(\frac{p_t}{m}q_o, \frac{m_t}{m}q_o, q_s,q_l\right)$. Additionally, we will denote the lab frame $\tau$, $\eta$, $\rho$, and $\varphi$ coordinates of $x^\mu\pm \frac{r^\mu}{2}$ as $\tau^\pm$, $\eta^\pm$, $\rho^\pm$, and $\varphi^\pm$, respectively.  Then, we isolate the $\q$ dependence of \eqref{s-bwm} as follows:
\begin{equation}
    S(\rr, \q;p^\mu) = \mathcal{N}\int dr^0 \int d^4x S_1\left( x^\mu+\frac{r^\mu}{2},p^\mu\right)S_1\left(x^\mu-\frac{r^\mu}{2},p^\mu\right) A_{CF}(x^\mu,r^\mu,\q;p^\mu)A_{T}(x^\mu,r^\mu,\q;p^\mu)
\end{equation}
where
\begin{multline}
    A^\mathrm{lab}_{CF}(x^\mu,r^\mu,\q;p^\mu) \\
    = \frac{\left[\left(m_t+\frac{p_t q_o}{m}\right)\cosh(\eta^+)-q_l\sinh(\eta^+)\right]\left[\left(m_t-\frac{p_t q_o}{m}\right)\cosh(\eta^-)+q_l\sinh(\eta^-)\right]}{m_t^2\cosh(\eta^+)\cosh(\eta^-)} \\
    = 1  - \frac{q_l}{m_t}\left[\tanh(\eta^+)-\tanh(\eta^-)\right] - \left(\frac{p_t q_o}{mm_t}\right)^2 \\
    - \frac{q_l^2}{m_t^2}\tanh(\eta^+)\tanh(\eta^-) + \frac{p_t q_o q_l}{mm_t^2} \left[\tanh(\eta^+)+\tanh(\eta^-)\right]
\end{multline}
and
\begin{multline}
    A^\mathrm{lab}_T(x^\mu,r^\mu,\q;p^\mu) = e^{-\frac{1}{T}q^\mathrm{lab}_\mu\left[u^\mu\left(x^\mu+\frac{r^\mu}{2}\right)-u^\mu\left(x^\mu-\frac{r^\mu}{2}\right)\right]} \\
    = 1 + \frac{1}{T}q_i\left[u'^i\left(x^\mu+\frac{r^\mu}{2}\right)-u'^i\left(x^\mu-\frac{r^\mu}{2}\right)\right]\\
    + \frac{1}{2T^2}q_iq_j \left[u'^i\left(x^\mu+\frac{r^\mu}{2}\right)-u'^i\left(x^\mu-\frac{r^\mu}{2}\right)\right]\left[u'^j\left(x^\mu+\frac{r^\mu}{2}\right)-u'^j\left(x^\mu-\frac{r^\mu}{2}\right)\right] \\ + \oo{q^3/T^3}
\end{multline}
with $u'^\mu$ being $u^\mu$ where the components are transformed to the PRF (but the arguments remain in the lab frame).

We note that both $A$ functions have no dependence on $\tau^\pm$. Additionally, all of the $\rho$ dependence in the source function and $A$ can be written as only a function of $\rho/R_0$. When we integrate $S(\rr,\q;p^\mu)$ (which gives us $1+q_i\ei + q_iq_j\eij + \oo{q^3/T^3}$), we can do a change of variables such that we integrate $d^4x^+d^4x^-$ in the $(\tau^\pm, \eta^\pm,\rho^\pm,\varphi^\pm)$ coordinates. Then the $d\tau^\pm$ integrals will factor out and cancel with $\mathcal{N}$, and a change of variables to $R^\pm = \rho^\pm/R_0$ will remove all $R_0$ dependence. This shows that the total size of the smoothness corrections as far as the source function is concerned (i.e. $\ei$ and $\eij$) is independent of the parameters $R_0$, $\tau_0$, and $\Delta\tau$. Those parameters solely affect the shape of the corrections $S^i$ and $S^{ij}$. This is important to keep in mind as we see the large effect of these parameters on the size of the smoothness corrections to the correlation function.

\subsection{On-Shell Corrections}
Since we are working with identical particles, $\kappa=1$ and $\alpha=\alpha_0=0$. Using $p^0=m_t$, we have
\begin{equation}
    P^0 = m_t + \frac{q^2}{2m_t} + \oo{q^3/m^2}.
\end{equation}
Now,
\begin{equation}\label{bwm-os}
    S(\rr,0;P^\mu) = \mathcal{N}\int dr^0 \int d^4x S_1\left( x^\mu+\frac{r^\mu}{2},p^\mu\right)S_1\left(x^\mu-\frac{r^\mu}{2},p^\mu\right) A_{CF}'(x^\mu,r^\mu;p^\mu)A_{T}'(x^\mu,r^\mu;p^\mu)
\end{equation}
where
\begin{equation}\label{aCFp}
    A_{CF}'(x^\mu,r^\mu;p^\mu) = \frac{(P^0)^2}{m_t^2} = 1 + \frac{q^2}{m_t^2} + \oo{q^3/m^3}
\end{equation}
and
\begin{multline}\label{aTp}
    A'{}^\mathrm{lab}_T(x^\mu,r^\mu,\q;p^\mu) = e^{-\frac{1}{T}(P^0-m_t)\left[u^0\left(x^\mu+\frac{r^\mu}{2}\right)+u^0\left(x^\mu-\frac{r^\mu}{2}\right)\right]} \\ 
    = 1 - \frac{q^2}{2Tm_t}\left[u^0\left(x^\mu+\frac{r^\mu}{2}\right)+u^0\left(x^\mu-\frac{r^\mu}{2}\right)\right]+ \oo{q^3/(mT^2)}.
\end{multline}
Because of the form of the on-shell expansion \eqref{os-expansion}, it is clear that a correction to the source function of the form $S'(\rr)=aS(\rr)$ will not contribute to the correlation function corrections at all, and thus the CF term does not generate nonzero on-shell corrections.

We get an additional contribution from $S'(\rr)$ from the frame transformation, which is the second correction in \eqref{sprime} (recalling that the first correction is zero). This correction is best computed numerically; in total we get
\begin{equation}
    q^2 S'(\rr;p^\mu) = S(\rr,0;P^\mu)-S(\rr,0;p^\mu) - \frac{p_t q^2}{2m^2m_t}\partial_{r_{out}}\int dr^0\, r^0 S(r^\mu,q^\mu;p^\mu)\mid_{q^0=0}
\end{equation}
where $S(\rr,0;P^\mu)-S(\rr,0;p^\mu)$ is found easily via \eqref{bwm-os}-\eqref{aTp}. Note that while the size of the frame transformation term will generally have $R$, $\tau_0$, and $\Delta\tau$ dependence, for the same reasons as the smoothness corrections the size of $S(\rr,0;P^\mu)-S(\rr,0;p^\mu)$ is independent of these parameters. As noted before, the frame transformation term does not contribute to $\varepsilon'$ anyways, so $\varepsilon'$ is independent of $R$, $\tau_0$, and $\Delta\tau$.

\section{Angle Average of an Odd-Order Kernel is Zero}\label{appC}

We will work in a single spin channel; the result then trivially holds for a sum of channels. We can write the wavefunction of \eqref{dwf} as
\begin{equation}
    \phi_{\q}(\rr) = \sum_{\ell} i^{\ell} (2\ell + 1) R_{q,\ell}(r) P_{\ell}\left(\frac{\q\cdot\rr}{qr}\right) = 4\pi\sum_{\ell,m} i^{\ell} R_{q,\ell}(r) Y_\ell^m(\rhat) Y_\ell^{m*}(\qhat).
\end{equation}
For ease, define $\rr_\pm = \rr \pm \frac{\zzeta}{2}$. Then,
\begin{equation}
    D_{\q} (\rr,\kk) = (4\pi)^2\int d^3\zzeta \, e^{-i\kk\cdot\zzeta} \sum_{\ell,\ell',m,m'}  i^{\ell-\ell'} R_{q,\ell}(r_+)R^*_{q,\ell'}(r_-) Y_\ell^m(\rhat_+)Y_{\ell'}^{m' *}(\rhat_-) Y_\ell^{m*}(\qhat)Y_{\ell'}^{m'}(\qhat).
\end{equation}
Now, using the orthonormality of the spherical harmonics we can take the angle average with respect to $\qhat$:
\begin{multline}
    \dbar_q(\rr,\kk) \equiv \idoqa D_{\q}(\rr,\kk) = 4\pi \int d^3\zzeta\, e^{-i\kk\cdot\zzeta} \sum_{\ell,m} R_{q,\ell}(r_+)R^*_{q,\ell}(r_-) Y_\ell^m(\rhat_+)Y_{\ell}^{m *}(\rhat_-) \\
    = \int d^3\zzeta\, e^{-i\kk\cdot\zzeta} \sum_{\ell} (2\ell+1) R_{q,\ell}(r_+)R^*_{q,\ell}(r_-) P_\ell(\rhat_+ \cdot \rhat_-).
\end{multline} 
But, for a real central potential, we know that $R_{q,\ell}(r)$ is real up to a constant complex phase $e^{i\delta_\ell}$. Therefore, $R_{q,\ell}(r_+)R^*_{q,\ell}(r_-)$ will be real and thus the $\kk$ dependence of $\dbar_q$ is the Fourier transform of a real function. This means $\dbar_q(\rr,-\kk)=\dbar_q(\rr,\kk)$, and thus all odd moments (in $\kk$) of $\dbar_q$ vanish. So,
\begin{equation}
    \idoqa \, K_{i_1, i_2...i_{2n+1}} = \int \frac{d^3\kk}{(2\pi)^3} k_{i_1} k_{i_2} \cdots k_{i_{2n+1}} \dbar_q(\rr,\kk) = 0.
\end{equation}

\end{document}